\documentclass[twocolumn]{aastex62}

\newcommand\xmm{{\em XMM-Newton}}
\newcommand\hst{{\em HST}}
\newcommand\swift{{\em Swift}}

\newcommand\grb{{sGRB\,160821B}}

\received{May 6, 2019}
\revised{July 20, 2019}
\accepted{August 4, 2019}
\submitjournal{ApJ}

\shorttitle{GRB\,160821B}
\shortauthors{Lamb et al.}

\begin{document}

\title{Short GRB 160821B: a reverse shock, a refreshed shock, and a well-sampled kilonova}

\correspondingauthor{G. P. Lamb}
\email{gpl6@leicester.ac.uk}

\author{G. P. Lamb}
\affiliation{University of Leicester, Department of Physics \& Astronomy and Leicester Institute of Space \& Earth Observation,  University Road,  
Leicester, LE1 7RH, UK}

\author{N. R. Tanvir}
\affiliation{University of Leicester, Department of Physics \& Astronomy and Leicester Institute of Space \& Earth Observation,  University Road,  
Leicester, LE1 7RH, UK}

\author{A. J. Levan}
\affiliation{Department of Physics, University of Warwick, Coventry, CV4 7AL, United Kingdom}
\affiliation{Department of Astrophysics, Radboud University, 6525 AJ Nijmegen, The Netherlands}

\author{A. de Ugarte Postigo}
\affiliation{Instituto de Astrof\'isica de Andaluc\'ia (IAA-CSIC), Glorieta de la Astronom\'ia s/n, 18008 Granada, Spain}
\affiliation{DARK, Niels Bohr Institute, University of Copenhagen, Lyngbyvej 2, 2100 Copenhagen \O,  Denmark}

\author{K. Kawaguchi}
\affiliation{Institute for Cosmic Ray Research, The University of Tokyo, 5-1-5 Kashiwanoha, Kashiwa, Chiba 277-8582, Japan}
\affiliation{Center for Gravitational Physics, Yukawa Institute for Theoretical physics, Kyoto University, Kyoto 606-8502, Japan}

\author{A. Corsi}
\affiliation{Department of Physics and Astronomy, Texas Tech University, Lubbock, TX 79409, USA}

\author{P. A. Evans}
\affiliation{University of Leicester, Department of Physics \& Astronomy and Leicester Institute of Space \& Earth Observation,  University Road,  
Leicester, LE1 7RH, UK}

\author{B. Gompertz}
\affiliation{Department of Physics, University of Warwick, Coventry, CV4 7AL, United Kingdom}

\author{D. B. Malesani}
\affiliation{The cosmic DAWN center, Niels Bohr Institute, University of Copenhagen, Lyngbyvej 2, 2100 Copenhagen \O, Denmark}
\affiliation{DARK, Niels Bohr Institute, University of Copenhagen, Lyngbyvej 2, 2100 Copenhagen \O, Denmark}

\author{K. L. Page}
\affiliation{University of Leicester, Department of Physics \& Astronomy and Leicester Institute of Space \& Earth Observation,  University Road,  
Leicester, LE1 7RH, UK}

\author{K. Wiersema}
\affiliation{Department of Physics, University of Warwick, Coventry, CV4 7AL, United Kingdom}
\affiliation{University of Leicester, Department of Physics \& Astronomy and Leicester Institute of Space \& Earth Observation,  University Road,  
Leicester, LE1 7RH, UK}

\author{S. Rosswog}
\affiliation{The Oskar Klein Centre, Department of Astronomy, AlbaNova, Stockholm University, SE-106 91 Stockholm, Sweden}

\author{M. Shibata}
\affiliation{Max Plank Institute for Gravitational Physics (Albert Einstein Institute), Am M\"uhlenberg 1, Potsdam-Golm, 14476, Germany}
\affiliation{Center for Gravitational Physics, Yukawa Institute for Theoretical physics, Kyoto University, Kyoto 606-8502, Japan}

\author{M. Tanaka}
\affiliation{Astronomical Institute, Tohoku University, Aoba, Sendai 980-8578, Japan}

\author{A. J. van der Horst}
\affiliation{Department of Physics, The George Washington University, 725 21st Street NW, Washington, DC 20052, USA}
\affiliation{ Astronomy, Physics, and Statistics Institute of Sciences (APSIS), 725 21st Street NW, Washington, DC 20052, USA}

\author{Z. Cano}
\affiliation{Berkshire College of Agriculture, Hall Place, Burchett's Green Road, Burchett's Green, Maidenhead, UK}

\author{J. P. U. Fynbo}
\affiliation{The cosmic DAWN center, Niels Bohr Institute, University of Copenhagen, Lyngbyvej 2, 2100 Copenhagen \O, Denmark}

\author{A. S. Fruchter}
\affiliation{Space Telescope Science Institute, 3700 San Martin Drive, Baltimore, MD 21218, USA}

\author{J. Greiner}
\affiliation{Max-Planck Institut f\"ur extraterrestrische Physik, D-85748 Garching, Giessenbachstr. 1, Germany}

\author{K. E.  Heintz}
\affiliation{Centre for Astrophysics and Cosmology, Science Institute, University of Iceland, Dunhagi 5, 107 Reykjav\'ik, Iceland}

\author{A. Higgins}
\affiliation{University of Leicester, Department of Physics \& Astronomy and Leicester Institute of Space \& Earth Observation,  University Road,  
Leicester, LE1 7RH, UK}

\author{J. Hjorth}
\affiliation{DARK, Niels Bohr Institute, University of Copenhagen, Lyngbyvej 2, 2100 Copenhagen \O, Denmark}

\author{L. Izzo}
\affiliation{Instituto de Astrof\'isica de Andaluc\'ia (IAA-CSIC), Glorieta de la Astronom\'ia s/n, 18008 Granada, Spain}

\author{P. Jakobsson}
\affiliation{Centre for Astrophysics and Cosmology, Science Institute, University of Iceland, Dunhagi 5, 107 Reykjav\'ik, Iceland}

\author{D. A. Kann}
\affiliation{Instituto de Astrof\'isica de Andaluc\'ia (IAA-CSIC), Glorieta de la Astronom\'ia s/n, 18008 Granada, Spain}

\author{P. T. O'Brien}
\affiliation{University of Leicester, Department of Physics \& Astronomy and Leicester Institute of Space \& Earth Observation,  University Road,  
Leicester, LE1 7RH, UK}

\author{D. A. Perley}
\affiliation{Astrophysics Research Institute, Liverpool John Moores University, IC2, Liverpool Science Park, 146 Brownlow Hill,  Liverpool L3 5RF, UK}

\author{E. Pian}
\affiliation{INAF, Astrophysics and Space Science Observatory, via P. Gobetti 101, 40129 Bologna, Italy}

\author{G. Pugliese}
\affiliation{Astronomical Institute Anton Pannekoek, University of Amsterdam,  PO Box 94249, 1090 GE Amsterdam, the Netherlands}

\author{R. L. C. Starling}
\affiliation{University of Leicester, Department of Physics \& Astronomy and Leicester Institute of Space \& Earth Observation,  University Road,  
Leicester, LE1 7RH, UK}

\author{C. C. Th\"one}
\affiliation{Instituto de Astrof\'isica de Andaluc\'ia (IAA-CSIC), Glorieta de la Astronom\'ia s/n, 18008 Granada, Spain}

\author{D. Watson}
\affiliation{The cosmic DAWN center, Niels Bohr Institute, University of Copenhagen, Lyngbyvej 2, 2100 Copenhagen \O, Denmark}

\author{R. A. M. J. Wijers}
\affiliation{Astronomical Institute Anton Pannekoek, University of Amsterdam,  PO Box 94249, 1090 GE Amsterdam, the Netherlands}

\author{D. Xu}
\affiliation{CAS Key Laboratory of Space Astronomy and Technology, National Astronomical Observatories, Chinese Academy of Sciences, Beijing 100012, China}



\begin{abstract}

We report our identification of the optical afterglow and host galaxy of the short-duration gamma-ray burst \grb. 
The spectroscopic redshift of the host is $z=0.162$, making it one of the lowest redshift sGRBs identified by \swift.
Our intensive follow-up campaign using a range of ground-based facilities as well as {\hst}, {\xmm} and {\swift}, shows evidence for a late-time excess of optical and near-infrared emission in addition to a complex afterglow.
The afterglow light-curve at X-ray frequencies reveals a narrow jet, $\theta_j\sim1.9^{+0.10}_{-0.03}$\,deg, that is refreshed at $>1$\,day post-burst by a slower outflow with significantly more energy than the initial outflow that produced the main GRB.
Observations of the 5\,GHz radio afterglow shows a reverse shock into a mildly magnetised shell.
The optical and near-infrared 
excess is fainter than AT2017gfo associated with GW170817, and is
well explained by
a kilonova with dynamic ejecta mass $M_{\rm dyn}=(1.0\pm0.6)\times10^{-3}$\,M$_{\odot}$ and a secular (postmerger) ejecta mass with $M_{\rm pm}=(1.0\pm0.6)\times10^{-2}$\,M$_\odot$, consistent with a binary neutron star merger resulting in a short-lived massive neutron star.
This optical and near-infrared dataset provides the best-sampled kilonova light-curve without a gravitational wave trigger to date.

\end{abstract}



\section{Introduction} \label{sec:intro}

Short-duration gamma-ray bursts (sGRBs) are widely thought to result from the merger of a binary neutron star (BNS) or a neutron star and a stellar mass black hole system. 
A fraction of the neutron star matter disrupted during the inspiral or collision will undergo rapid accretion onto the remnant object and launch an ultra-relativistic jet \citep[e.g.][]{nakar07,gehrels09}.
Energy dissipation within such a jet produces a GRB, and, as this outflow decelerates, an external shock forms producing broad-band afterglow emission.
This progenitor model is supported by the fact that well-localised sGRBs (mainly the sample discovered by the {\em Neil Gehrels Swift Observatory}, hereafter referred to as {\em Swift}) appear to be produced in a wide range of stellar populations, including those with no recent star formation, and on occasions at large distances (10s of kpc in projection) from their putative host galaxies \citep[e.g.][]{fong13,tunnicliffe14}.

A further signature of compact binary mergers involving neutron stars is via the observation of a slower transient, variously called a `macronova' \citep{kulkarni2005}, `kilonova' \citep{metzger2010}, or `merger-nova' \citep{gao2015} (in this paper we shall use the term kilonova).
A kilonova is powered by the radioactive decay of heavy, unstable, neutron-rich species created from decompressed neutron star material which is ejected during the merger \citep[e.g.][]{li98}.

The first compelling observational evidence for such a kilonova was the case of sGRB\,130603B, for which excess near-infrared emission was detected in {\em Hubble Space Telescope} (\hst) imaging at about one week in the rest frame after the event \citep{tanvir13,berger13}.
That this excess appeared in the near-IR tallied with predictions that the
same heavy r-process elements created in the kilonova should produce dense line-blanketing in the optical,
leading to emission appearing in the near-IR in the days to weeks following the merger \citep{barnes13, kasen13, tanaka13}.
A further interest in these events comes from the fact that this process of radioactive decay naturally leads to stable r-process elements, thus potentially explaining the abundances of more than half the elements in the universe heavier than iron \citep[e.g.][]{lattimer74,freiburghaus99,rosswog18}.  
Mapping the diversity and evolution of kilonova events over cosmic time is therefore an essential ingredient to quantifying their global contribution to nucleosynthesis.

At a redshift $z=0.36$ \citep{deUP14}, identifying the kilonova emission in the afterglow to sGRBs\,130603B was challenging and would not currently be feasible at higher redshifts, where the bulk of well-localised sGRBs have been found. 
Indeed, state-of-the-art modelling of neutron-star binary mergers suggests that ejection of sufficient material to create a kilonova as bright as this is unlikely to happen in most mergers, and may require special circumstances such as a high mass-ratio for the components of the binary \citep[e.g.][]{hotoke2013, just2015, sekiguchi2016}.
Nonetheless, following this discovery, and based on archival data, possible kilonova signatures were identified via a late-time $I$-band excess emission in two earlier GRBs;
namely sGRB\,050709 at $z=0.16$ \citep{jin16}, and GRB\,060614 at $z=0.125$ \citep{yang15}.
More recently,
{it has been proposed that
the optical counterparts identified
for sGRB 070809 at $z=0.22$ \citep[][although note that the host identification, and therefore redshift, in this case is rather uncertain]{jin2019} and sGRB 150101B at $z=0.13$ \citep{troja2018} may have been dominated by kilonova emission}.
For GRB\,060614 the claim is particularly controversial in that its prompt duration, $T_{90}\sim100$\,s, is much longer than the canonical $T_{90}\leq2$\,s for a sGRB.
However, the absence of an accompanying bright supernova combined with it exhibiting an initial spike of gamma-rays with duration of only a few seconds has led to speculation that it could have been produced by a compact binary merger \citep{galyam2006, gehrels2006, perley2009, kann2011}.

The recent multi-messenger observation of the BNS merger GW170817, discovered via gravitational waves and associated with a burst of $\gamma$-rays, GRB\,170817A, detected by {\em Fermi} and {\em INTEGRAL} \citep{abbott17a, abbott17b, goldstein2017, savchenko2017}, provided an opportunity to test directly the merger progenitor model.
GRB\,170817A appeared faint when compared to the cosmological sample of sGRBs and by considering the compactness problem and lack of an early afterglow indicates that the burst of $\gamma$-rays is unlikely to be a typical sGRB seen off-axis \citep[e.g.][]{lamb2018a, ziaeepour2018, matsumoto2019}, however, \cite{ioka2019} show that the observed GRB emission likely originates from a `mid'-region of a structured outflow.
The rapid decline and super-luminal motion of the late-time afterglow to GW170817 offer strong support for the sGRB - BNS association \citep{ghirlanda2018, mooley2018, vaneerten2018, lamb19}.
Additionally, a kilonova was seen to follow GW170817, and monitored intensively at UV, optical and near-infrared wavelengths \citep[e.g.][]{coulter17,tanvir17,evans17,smartt17,pian17,andreoni17,kasliwal17a,cowperthwaite17}.
By scaling the well-sampled GW170817 kilonova lightcurve to the distance of sGRBs with afterglows, attempts have been made to investigate the diversity of the kilonova population \citep{gompertz18, ascenzi2018, rossi2019}.

Here we report a search with \hst, \xmm, and ground-based telescopes including the Gran Telescopio Canarias (GTC), the Nordic Optical Telescope (NOT), the Telescopio Nazionale Galileo (TNG), the William Herschel Telescope (WHT), and the Karl G. Jansky Very Large Array (VLA) for afterglow and kilonova emission accompanying \grb, associated with a morphologically disturbed host galaxy at $z = 0.162$. 
We supplement these data with publicly available and/or published in other sources \textit{Swift}, VLA, and \textit{Keck} data.
Throughout we assume a flat universe with $\Omega_{\rm m} = 0.308$ and $H_{\rm 0} = 67.8$\,km\,s$^{-1}$\,Mpc$^{-1}$ \citep{planck15}. 
Optical and near-IR magnitudes are reported on the AB system.
In \S \ref{sec:obs} we report the observations at X-ray, optical, near-IR, and radio frequencies plus the identification of the afterglow and the host.
The results, interpretation and afterglow and kilonova modelling are shown in \S \ref{sec:results}.
We discuss these results in \S \ref{sec:discussion} and give concluding remarks in \S \ref{sec:conclusions}.

\section{Observations}
\label{sec:obs}

\subsection{Discovery of \grb}

The  Burst Alert Telescope (BAT) onboard {\em Swift} triggered on \grb{} on 2016 Aug 21 at 22:29 UT.
The reported duration of the burst was $T_{90}(15-350\,{\rm keV})=0.48\pm0.07$ s \citep{palmer16gcn}.
The burst was also detected by {\em Fermi}/GBM, from which a somewhat longer duration of
$\approx1$\,s was found \citep{stanbro16gcn}.
\citet{lu17} performed a joint fit to the \swift/BAT and {\em Fermi}/GBM data, finding the total fluence  in the 8-10000\,keV band of $(2.52 \pm 0.19) \times 10^{-6}$\,erg\,cm$^{-2}$. This corresponds to an isotropic energy, assuming the redshift of $z=0.162$, of $E_{\gamma,\rm iso}=(2.1\pm0.2)\times10^{50}$\,erg, fairly typical of the population of short GRBs with measured redshifts \citep{berger14}.





\subsection{Afterglow identification}
\label{sec:afterglow}

After slewing, the X-ray Telescope (XRT) on \swift\ detected a fading afterglow which provided a
refined localisation, and from the X-ray spectrum found no evidence for
significant absorption beyond that expected due to foreground gas in our Galaxy
\citep{sbarufatti16gcn}.
As described below, our early optical imaging  identified the afterglow of the burst and  a prominent nearby galaxy 
at a separation of about 5.7 arcsec \citep{xu16gcn}.

With a magnitude of $r\approx19.4$ (Section~\ref{sec:host}), the probability of the chance alignment of an unrelated galaxy of this brightness or brighter this close to the line of sight is $P_{\rm chance}\approx1.5$\% \citep[using the formalism of][]{bloom02} and although low, is not entirely negligible.
However, the absence of any faint underlying quiescent emission in our final \hst\ epochs (see Section~\ref{sec:afterglow}), which
might otherwise suggest a higher redshift host, adds support to our working hypothesis that this is the host galaxy 
of \grb.

The Nordic Optical Telescope, located in the Canary Islands (Spain), began optical observations at 23:02 UT, only 33 minutes post-burst.
These revealed an uncatalogued point source within the X-ray error region, presumed
to be the optical afterglow \citep{xu16gcn}. The best astrometry came 
from our \hst{} images, and gave a position of
 RA(J2000) = 18:39:54.550,
Dec(J2000) = +62:23:30.35
with an uncertainty of $\approx0.03$ arcsec in each coordinate, registered on the GAIA DR2 astrometric reference frame \citep{gaia2016,gaia2018}.
\citet{fong16gcn} reported a detection of the radio afterglow at 5\,GHz with the VLA, which provided
a burst location of
RA(J2000) = 18:39:54.56,
Dec(J2000) = +62:23:30.3
(reported error 0.3 arcsec), consistent with our \hst\ localisation.

\subsection{Host galaxy and redshift}
\label{sec:host}

\begin{figure*}[ht!]
\plotone{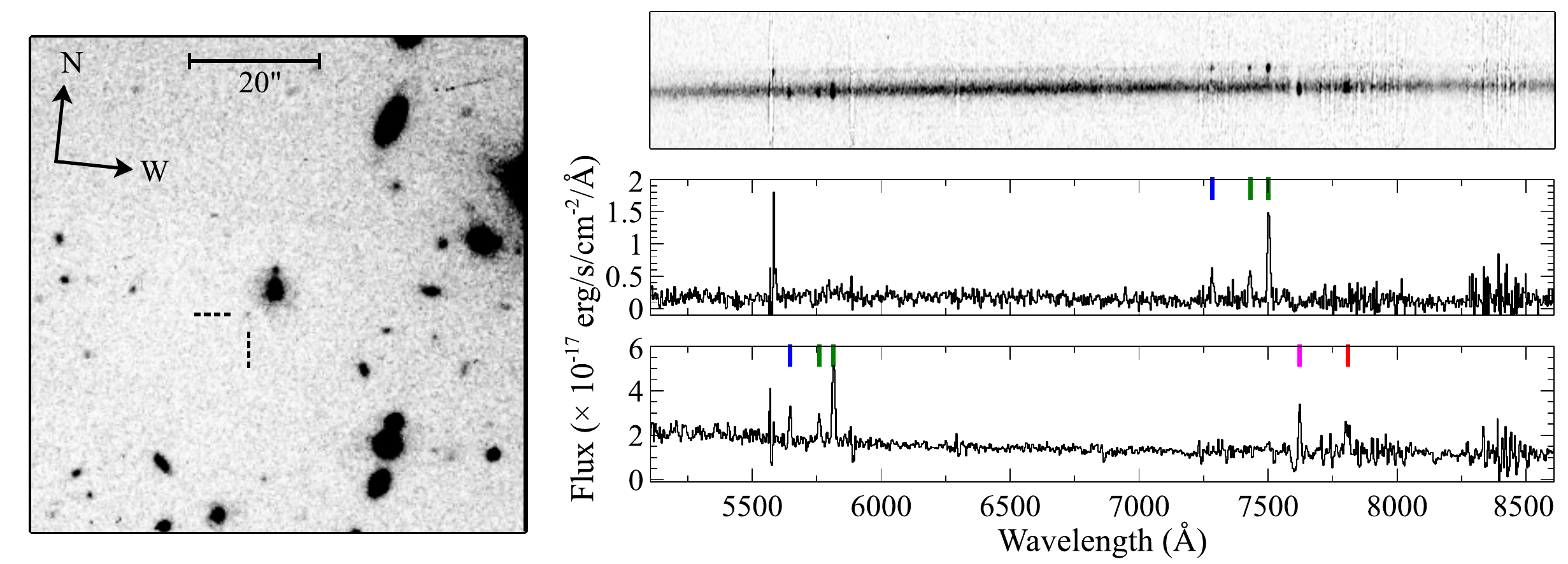}
\caption{{Left: WHT ACAM $z$-band image from 1.08 days post burst, see Table \ref{tab:oir}. The transient location is indicated by the dashed lines. Right panels:} The spectrum obtained with WHT/ACAM of the putative host galaxy at $z=0.1616$ (brighter, lower trace showing prominent lines of H$\alpha$, H$\beta$, [\ion{S}{2}] and [\ion{O}{3}] {indicated with short vertical lines in blue, pink, red and green respectively}) and
a presumably unrelated background galaxy at $z=0.4985$ (fainter, upper trace).
\label{fig:whtspec}}
\end{figure*}

\begin{figure*}[ht!]
\plotone{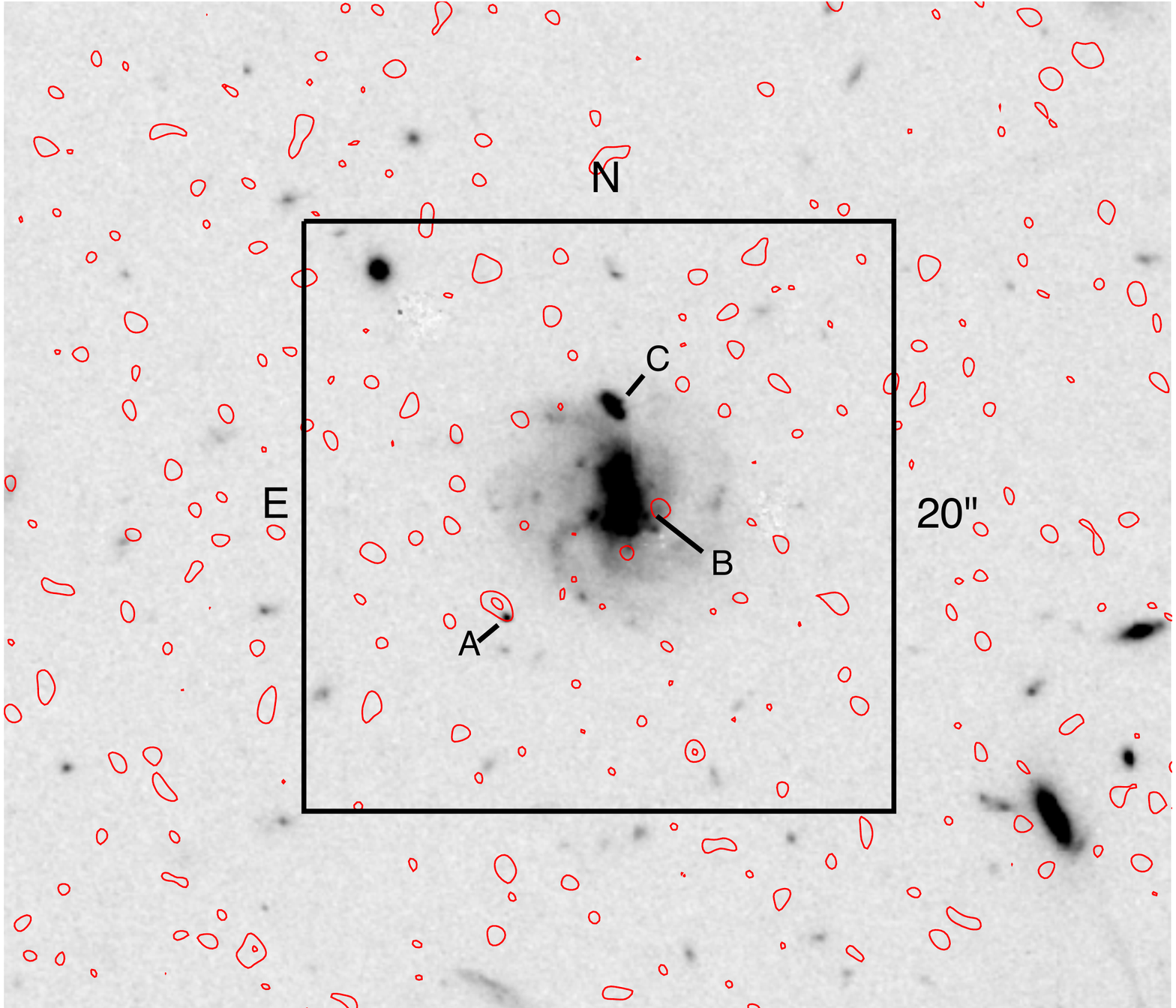}
\caption{The epoch 1 {(3.7\,days post-burst)} F110W+F160W \hst{} image of the field of  \grb, showing (A) the near-IR counterpart of
the burst, (B) the proposed host galaxy at $z=0.162$, (C) a background galaxy at $z=0.5$.
{The red contours show 1.5$\sigma$ and 3$\sigma$ radio flux increments at $\sim10$\,days post burst. The slight spatial offset of the radio and optical sources is consistent with the effects of noise in the map.}
\label{fig:field}}
\end{figure*}

The position of the proposed host galaxy measured from our \hst\ images is
RA(J2000) = 18:39:53.968,
Dec(J2000) = +62:23:34.35.
We obtained spectroscopy of this galaxy with the WHT
using the Auxiliary Port Camera (ACAM), in observations beginning on 2016 Aug 22 at 22:57 UT  \citep{levan16gcn}. 
The data were reduced using standard {\sc IRAF} routines.
The resulting 2D and 1D extracted spectra are shown in Figure~\ref{fig:whtspec}, with emission 
lines of H$\alpha$, H$\beta$, [\ion{S}{2}] and [\ion{O}{3}] providing a redshift of $z=0.1616\pm0.0002$.
The slit was aligned to cross both the nucleus of the main galaxy and a fainter blob of emission
to the north, labelled `B' and `C' respectively on Figure~\ref{fig:field}. The latter turned out to be a higher redshift galaxy\footnote{For completeness, we note that the impact parameter of the GRB from this background galaxy is $\approx50$\,kpc, and it has a $P_{\rm chance}\approx40\%$, confirming that it is not a good alternative host candidate.}
at $z=0.4985 \pm0.0002$,
the spectrum of which is also shown in Figure~\ref{fig:whtspec}.


At a redshift $z=0.162$ the separation between afterglow  and host corresponds to 16.4\,kpc in projection, which
is consistent with the offset distribution found for other sGRBs \citep{fong13a,tunnicliffe14}.

Morphologically,  the host appears to be a face-on, disturbed spiral galaxy (Figure~\ref{fig:field}). 
The extended, warped appearance of the central bulge suggests an ongoing merger, and the nebular emission lines
are consistent with active star formation.
It is interesting to note, although most likely coincidental, that the hosts of both sGRB\,130603B and GRB\,170817A were also notably disturbed \citep{tanvir13,levan17}.

The foreground extinction corrected magnitude of the host from the \hst{} imaging (with the flux from the $z=0.5$ background
galaxy subtracted) is $r_{606,0}=19.4$. This corresponds to an absolute magnitude of $M_r=-20.0$, which is 
$\sim L^*/3$ with respect to the \citet{loveday15} ``blue'' (star-forming) galaxy population.

The $r$-band  25 mag arcsec$^{-2}$ isophote has a radius of  $\approx3.5$\,arcsec, corresponding to a linear scale of $\approx 10$\,kpc.
However it is possible to trace lower surface brightness emission from the galaxy out to the GRB location, albeit at a faint surface
brightness level of $\approx 27$ $r$ mag arcsec$^{-2}$.

\subsection{Further optical and near-infrared monitoring}
\label{sec:oir}

\grb{} is amongst the lowest redshift sGRBs found by \swift{} to date. This, combined with its comparatively low foreground Galactic extinction of $A_V = 0.118$ mag \citep{schlafly11}, motivated an intensive follow-up monitoring campaign.

Further optical and near-IR imaging was obtained with the NOT, the GTC, and the WHT
over the next several nights.
These data were reduced using standard procedures, and calibrated photometrically using Pan-STARRS (optical) and 2MASS (near-IR) stars in the field.

Observations with the \hst{} using the Wide Field Camera 3 (WFC3), were obtained in
the F606W filter (a wide filter spanning approximately the $V$ and $r$ bands), the F110W filter (a wide $YJ$ band) and the F160W filter
($H$ band) from several days to several weeks post-burst \citep{troja16}. We adopted the standard photometric calibration for these bands\footnote{\url{http://www.stsci.edu/hst/wfc3/phot_zp_lbn}}, and aperture corrections were determined using bright point sources on the frames.

In all cases, interactive aperture photometry was performed using the Gaia software\footnote{
\url{http://astro.dur.ac.uk/~pdraper/gaia/gaia.html}}.
Care was taken to obtain sky estimates close to the position of the transient, since the background was not entirely free of light from the host galaxy.

These observations revealed the counterpart to be initially steady in brightness during the observations made on the first night, but thereafter it faded monotonically in all bands.
In the third {\em HST} visit, at $\sim23$\,days, no emission is detected at the burst location, which was confirmed by a final visit at $\approx$100\,days.
A summary of the results of all our optical and near-IR photometry for the \grb{} afterglow, together with selected magnitudes reported elsewhere, is presented in Table~\ref{tab:oir}.

\begin{deluxetable*}{cClcCl}[htb!]
\tablecaption{Optical and near-IR photometry of the \grb{} afterglow \label{tab:oir}}
\tablecolumns{6}
\tablenum{1}
\tablewidth{0pt}
\tablehead{
\colhead{$\Delta t$ (d)} & \colhead{$t_{\rm exp}$ (s)} & \colhead{Telescope/Camera} & \colhead{Filter} &  \colhead{{$AB_{\rm 0}$}} & \colhead{Source of photometry} }
\startdata
0.95 & 14\times300& TNG/DOLoRes & $g$ & 24.02\pm0.16 & This work \\ %
2.02 & 7\times120 &  GTC/OSIRIS & $g$ & 25.56\pm0.16 & This work \\
3.98 & 10\times120 & GTC/OSIRIS & $g$ & 25.98\pm0.15 & This work \\
6.98 & 21\times120 &  GTC/OSIRIS & $g$ & 26.90\pm0.18 &  This work \\
0.05 & 6\times300 & NOT/AlFOSC  & $r$ & 22.58\pm0.09 & This work \\
0.07 & 6\times300 & NOT/AlFOSC  & $r$ & 22.52\pm0.06 & This work \\
0.08 & 3\times90 & GTC/OSIRIS  & $r$ & 22.53\pm0.03 & This work \\
1.06 & 6\times240 & WHT/ACAM & $r$ & 23.82\pm0.07 & This work \\
1.95 & 9\times300 & NOT/AlFOSC & $r$ & 24.81\pm0.07 & This work \\
2.03 & 5\times120 & GTC/OSIRIS & $r$ & 24.80\pm0.06 & This work \\
3.64 & 4\times621 & HST/WFC3/UVIS & F606W & 25.90\pm0.06 & This work \\
4.99 & 27\times120 & GTC/OSIRIS & $r$ & 26.12\pm0.25 & This work \\
10.40 & 4\times621 & HST/WFC3/UVIS & F606W & 27.55\pm0.11 & This work \\
23.20 & 1350 & HST/WFC3/UVIS & F606W &  >27.34 & This work  \\
0.08 & 3\times90 & GTC/OSIRIS & $i$ & 22.37\pm0.03 & This work \\
2.04 & 5\times90 & GTC/OSIRIS & $i$ & 24.44\pm0.10 & This work \\
4.00 & 3\times90 &  GTC/OSIRIS & $i$ & 25.70\pm0.38 &  This work \\
9.97 & 18\times90 &  GTC/OSIRIS & $i$ & >25.59 &  This work \\
0.08 & 3\times60 & GTC/OSIRIS & $z$ & 22.39\pm0.02 & This work \\
1.08 & 6\times240 & WHT/ACAM & $z$ & 23.60\pm0.15 & This work \\
1.99 & 9\times300 & NOT/AlFOSC & $z$ & 23.90\pm0.23 & This work \\
2.04 & 7\times60 & GTC/OSIRIS  & $z$ & 24.34\pm0.24 & This work \\
3.76 & 2397 & HST/WFC3/IR & F110W  & 24.69\pm0.02 & This work \\
10.53 & 2397 & HST/WFC3/IR & F110W & 26.69\pm0.15 & This work \\
23.18 & 1498 & HST/WFC3/IR & F110W &  >27.34 & This work  \\
0.96 & 33\times20 & GTC/CIRCE & $H$  & 23.83\pm0.35 & This work \\
3.71 & 2397 & HST/WFC3/IR & F160W & 24.43\pm0.03 & This work \\
10.46 & 2397 & HST/WFC3/IR & F160W & 26.55\pm0.23 & This work \\
23.23 & 2098 & HST/WFC3/IR & F160W & >27.21 & This work  \\
4.3 & 45\times30.8 & Keck/MOSFIRE & $K$ & 24.04^{+0.44}_{-0.31} & \citet{kasliwal17b} \\
\enddata
\tablecomments{Column (1) mid-time of observation with respect to GRB trigger time. Magnitudes corrected for Galactic foreground extinction
according to $A_V=0.118$ from \citet{schlafly11}.}
\end{deluxetable*}

\subsection{X-ray monitoring}
\label{sec:xray}

\swift/XRT monitoring continued for 
2.5 days, showing evidence for a significant break to a steeper rate of fading around 0.4 days.
Our {\xmm} observations comprised two visits at approximately 4 and 10 days post-burst.
The first visit produced a very significant detection, and was above a simple extrapolation between the last  {\swift} visits. This is discussed further in \S \ref{sec:results}.

A summary of the X-ray observations is presented in Table~\ref{tab:x}.

\subsection{Radio monitoring}

The 5\,GHz radio detection in 1 hour of observations at 3.6 hours after the burst had a reported flux density of $\sim35$\,$\mu$Jy;
an additional observation with the same telescope at 26.5 hours post-burst returned a 3$\sigma$ upper limit of 18 $\mu$Jy \citep{fong16gcn}.

Late-time radio observations of the GRB\,160821B field were carried out with the VLA, at a central frequency of about 10 GHz and nominal bandwidth of 4 GHz. 
The first observation started on 2016 September 01 at 23:24:16 UT; 
the second observation started on 2016 September 08 at 00:10:33 UT. 
Data were calibrated using the automated VLA calibration pipeline available in the Common Astronomy Software Applications (CASA). 
After calibration, data were inspected for flagging, and then imaged using the CLEAN algorithm available in CASA. 
For each of the observations, we estimated the maximum flux density measured within a circular region centered around the position of GRB\,160821B and with a radius of 0.6 arcsec (comparable to the nominal FWHM of the VLA synthesized beam in its B configuration at 10 GHz). 
If the maximum peak density found within this region is above $3\times$ the image rms, then we report the measured flux density value and assign to it an error obtained by adding in quadrature the image rms and a 5\% absolute flux calibration error. 
On the other hand, if the maximum flux density within the selected circular region does not exceed the $3\times$ rms, we report an upper-limit with value equal to $3\times$ the image rms.
Radio data\footnote{{We note that the measured radio flux at $\sim17$\,days is $\sim31~\mu$Jy and only just below $3\times$ the image rms. The presented upper-limit at this time, $<33~\mu$Jy, is likely an underestimate, where the flux at the GRB location plus $2\sigma$ would give a limit of $<53~\mu$Jy.}} are listed in Table \ref{tab:r}.

\begin{deluxetable}{cc}[htb!]
\tablecaption{{\swift} (top) and {\xmm} (bottom) X-ray observations in the 0.3--10\,keV band, of the \grb\ afterglow after the first hour. \label{tab:x}}
\tablecolumns{2}
\tablenum{2}
\tablewidth{\textwidth}
\tablehead{
\colhead{$t$}  & \colhead{0.3-10 keV flux} \\
 (d)& ($10^{-14}$ erg cm$^{-2}$ s$^{-1}$) }
\startdata
$0.06^{+0.01}_{-0.01}$ & $59.6^{+10.8}_{-10.8}$ \\
$0.14^{+0.06}_{-0.02}$ & $45.8^{+7.50}_{-7.50}$ \\
${ 0.30^{+0.03}_{-0.03}}$ & ${ 32.1^{+9.46}_{-7.45}}$ \\
${ 0.34^{+0.01}_{-0.01}}$ & ${ 28.0^{+7.42}_{-6.00}}$ \\
${ 0.42^{+0.13}_{-0.02}}$ & ${ 13.1^{+3.74}_{-2.99}}$ \\
${ 1.02^{+0.39}_{-0.30}}$ & ${ 3.44^{+1.49}_{-1.10}}$ \\ 
${ 2.33^{+2.11}_{-0.67}}$ & ${ \leq2.53}$ \\ \hline
$3.91^{+0.12}_{-0.12}$ & $1.70^{+0.21}_{-0.21}$ \\
$9.95^{+0.17}_{-0.17}$ & $0.51^{+0.20}_{-0.20}$
\enddata
\tablecomments{Column (1) - times of observation with respect to GRB trigger time, uncertainties represent the duration of the observation; column (2) - fluxes corrected for Galactic foreground absorption following the prescription of \citet{willingale13}.}
\end{deluxetable}

\begin{deluxetable}{cccl}[htb!]
\tablecaption{Radio data used in the analysis \label{tab:r}}
\tablecolumns{4}
\tablenum{3}
\tablewidth{0pt}
\tablehead{
\colhead{$t$} & \colhead{$\nu$}  & \colhead{Flux density} & \colhead{Source} \\
(d) & (GHz) & (mJy) &}
\startdata
$0.15$ & $5.0$ & $0.035$ & \cite{fong16gcn} \\
$1.10$ & $5.0$ & $<0.018$ & \cite{fong16gcn} \\
$10.06$ & $9.8$ & $0.016\pm0.004$ & This work \\
$17.09$ & $9.8$ & $<0.033$ & This work
\enddata
\tablecomments{Column (1) - times of observation with respect to GRB trigger time. Column (2) - central frequency. Column (3) - Flux density. Column (4) - source, where `This work' refers to observations by the VLA in B configuration under program VLA/16B-386 (PI: Gompertz).}
\end{deluxetable}

\section{Light-curve Behaviour, Interpretation, and Modelling}
\label{sec:results}

In this section we describe the behaviour of the light-curve at the various observed frequencies.
Additionally, we give our interpretation of this behaviour before estimating the light-curve with physically motivated models.
These models provide 
parameter estimates for the various contributing emission components. 

\subsection{X-ray frequency light-curve behaviour}
\label{sec:xray}

\begin{figure}[ht!]
\plotone{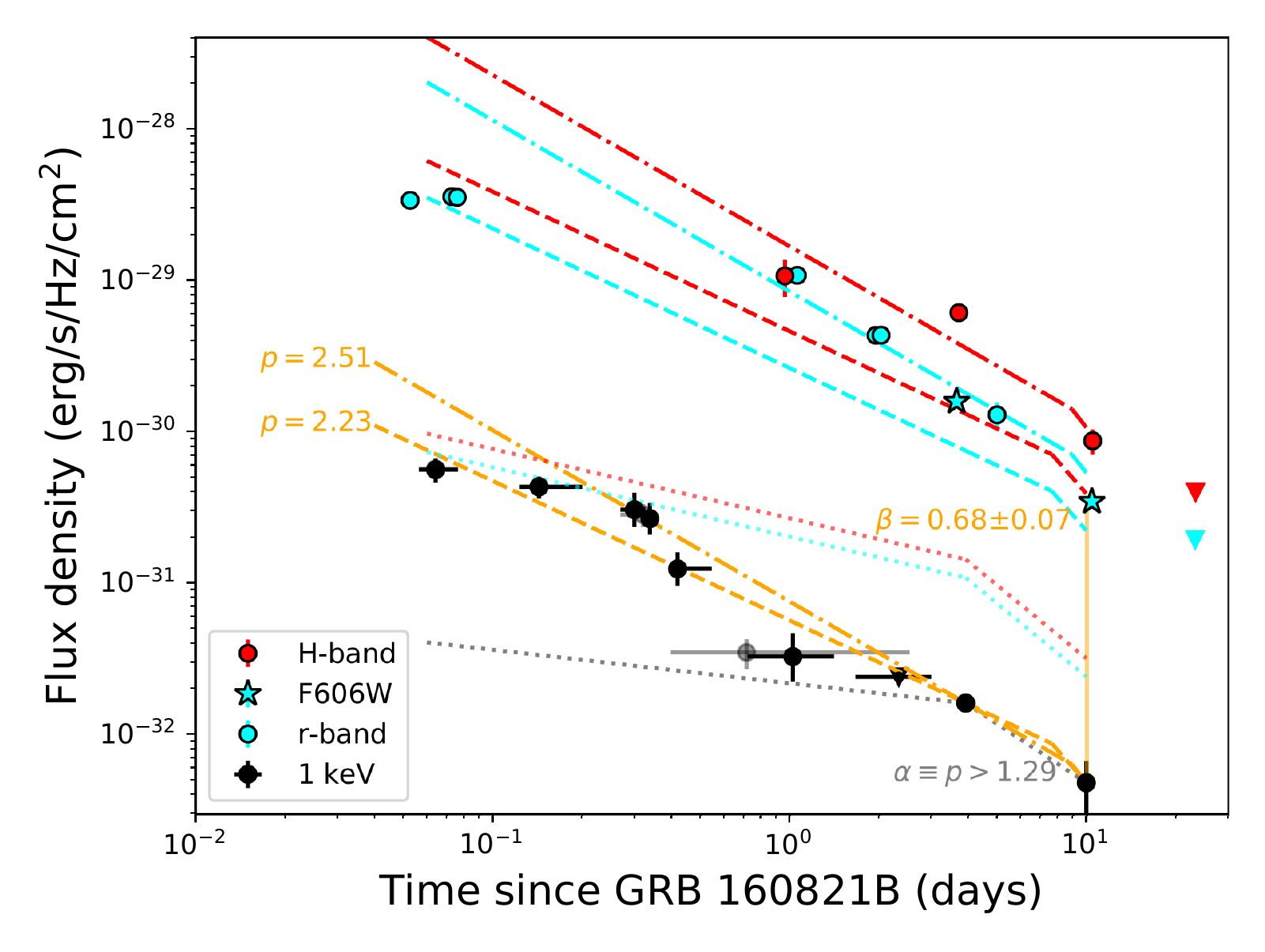}
\caption{Light curves of the \grb{} afterglow. The X-ray data point horizontal bars represent the duration of the observations, and therefore are not error-bars. \swift/XRT data from 0.1 to 3 days are re-binned to highlight the steep decline at $\gtrsim 0.3$ days and the low count rate at $\sim 2$--3 days (original binned data are shown as grey symbols);  black markers show a detection with associated uncertainty and triangles indicate upper limits. Dashed and dash-dotted orange lines, representing the limits on a simple power-law afterglow, consistent with the spectral gap between the X-ray and the $r$-band data at 10 days are shown, see \S \ref{sec:results} for details {(we plot F606W data (star symbol) as $r$-band)}. A jet break at $\sim7$ days is required when assuming this temporal behaviour. The $r$- and $H$-band optical data are shown in cyan and red, respectively, with a power-law light-curve extrapolated from the dashed/dash-dotted X-ray limits. Ignoring the optical to X-ray spectral constraints, the minimum power-law permitted by the late X-ray data, i.e. assuming a jet-break at $\sim4$ days, is shown as a dotted grey line at 1 keV and extrapolated to the expected $r$- and $H$-bands afterglow in cyan and red.
\label{fig:lcx}}
\end{figure}

A period of extended emission\footnote{Due to the lack of a clear or consistent definition for extended emission in GRBs, we follow \cite{kisaka15} who define extended emission as X-ray emission with a duration $\sim10^2$\,s and indicative of a long-lasting central engine. We additionally note that \grb{} is included in the sample of sGRBs with EE by \cite{kisaka2017} and \cite{kagawa2019}.} (EE) follows the \grb{} prompt emission for a duration of $\sim 200$--300\,s.
Following the rapid decline of the EE, \swift/XRT and \xmm{} observations show a shallower decline between $\sim0.01$ and 10~days;
as expected from an afterglow.
However, this late-time X-ray flux deviates from the expected power-law decline of a simple  afterglow model.
The flux level drops below that expected from a power-law decay between $\sim 0.3$ to 4 days. 
Re-binning the \swift/XRT data into photon bins with a lower minimum count, the behaviour of the X-ray light-curve is more clearly revealed; 
see Figure~\ref{fig:lcx} where the grey markers show the data using the typical minimum photon count per bin and the black markers show the re-binned flux levels (a triangle indicates an upper-limit).
A photon index $\Gamma=1.7$ is assumed, which is consistent with both \swift/XRT ($\Gamma=2.0^{+0.7}_{-0.6}$) and \xmm{} ($\Gamma=1.4^{+0.5}_{-0.4}$).
Horizontal error-bars indicate the duration of the observations at each point. 
The re-binned data reveal a break in the X-ray light-curve at $\sim0.35$ days{, where the flux drops significantly for all the following data,} and the flux level at 2--3 days is comparable to the \xmm{} observed flux level at $\sim 4$ days.

\subsection{Behaviour at optical and near-infrared frequencies}

\begin{figure}[ht!]
\plotone{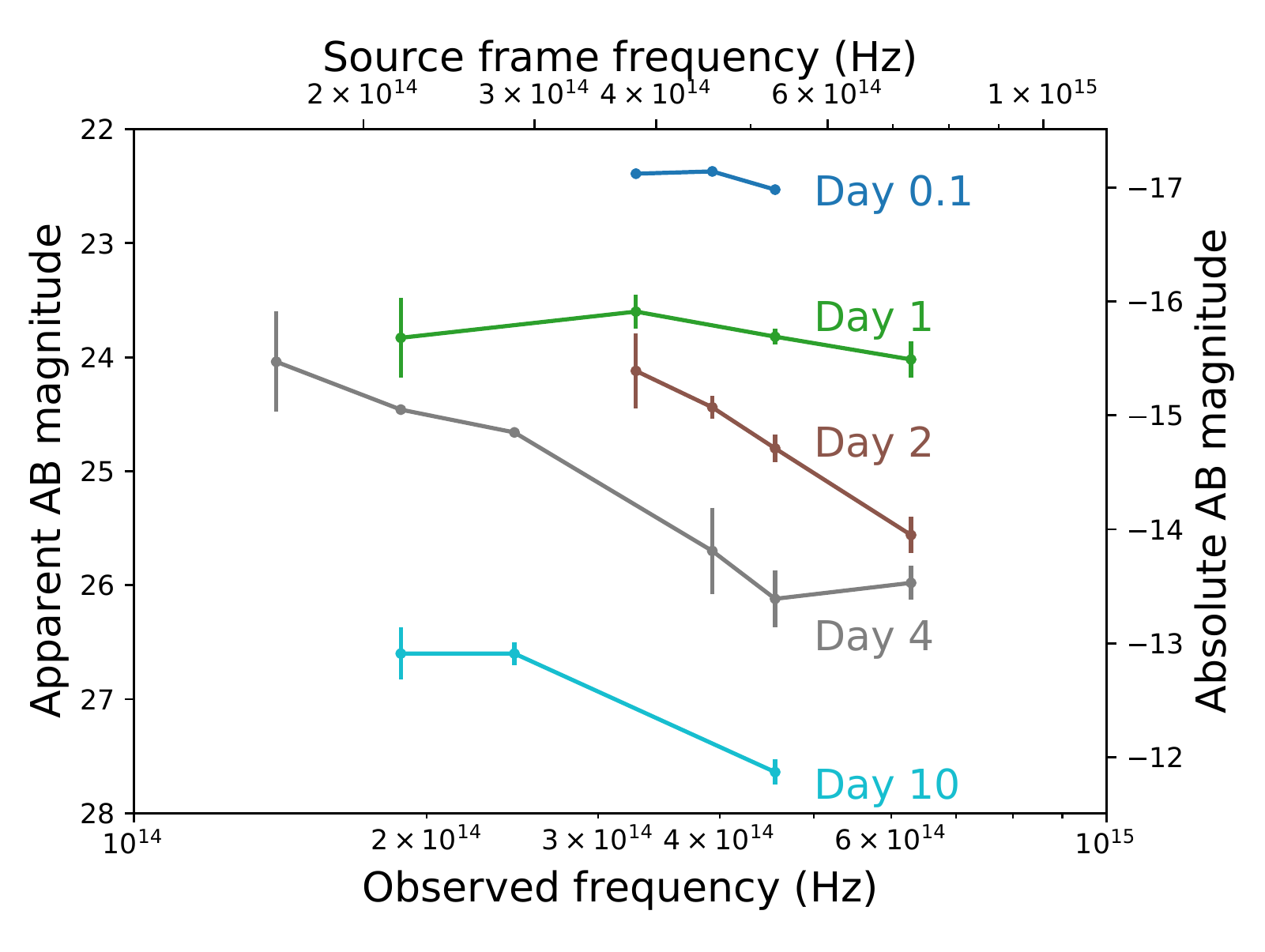}
\caption{The spectral energy distribution of the transient at five epochs, illustrating the large changes in colour, from blue to red. The photometry has been corrected for foreground Galactic extinction.
\label{fig:sed}}
\end{figure}

Figure~\ref{fig:sed} shows the spectral energy distribution of all the optical data from Table~\ref{tab:oir}, where we have averaged together points taken in the same filter at close to the same time.
The colour evolution of the transient exhibits a trend from blue in observations taken roughly one day after the burst to  a much redder colour in all subsequent detections. 
This is immediately indicative of an emerging kilonova component which itself is evolving from blue to red on time-scales of days; \cite[see e.g.][]{perego2014, tanaka2018, wollaeger2018}.
$r$- and $H$-band data are shown in Figure~\ref{fig:lcx} for comparison with a typical power-law decline extrapolated from the power-law used to show the behaviour at X-ray frequencies (see \S\ref{sec:xray}).
The deviation from a power-law with an excess in blue and then red is evident;
the behaviour at optical and near-IR is distinct from that at 1 keV.

{We note that while treating the F606W magnitudes as $r$-band in principle introduces a systematic error, the measured $g$-F606W colour is flat (consistent with our interpretation below that the optical light is afterglow dominated at these times),
indicating that colour corrections would be smaller than the photometric errors. (Furthermore, even for our kilonova models, at the time of those epochs, the predicted difference between F606W and the $r$-band is $\lesssim 0.2$\,AB mag.) }

\begin{figure*}
\includegraphics[width=20cm]{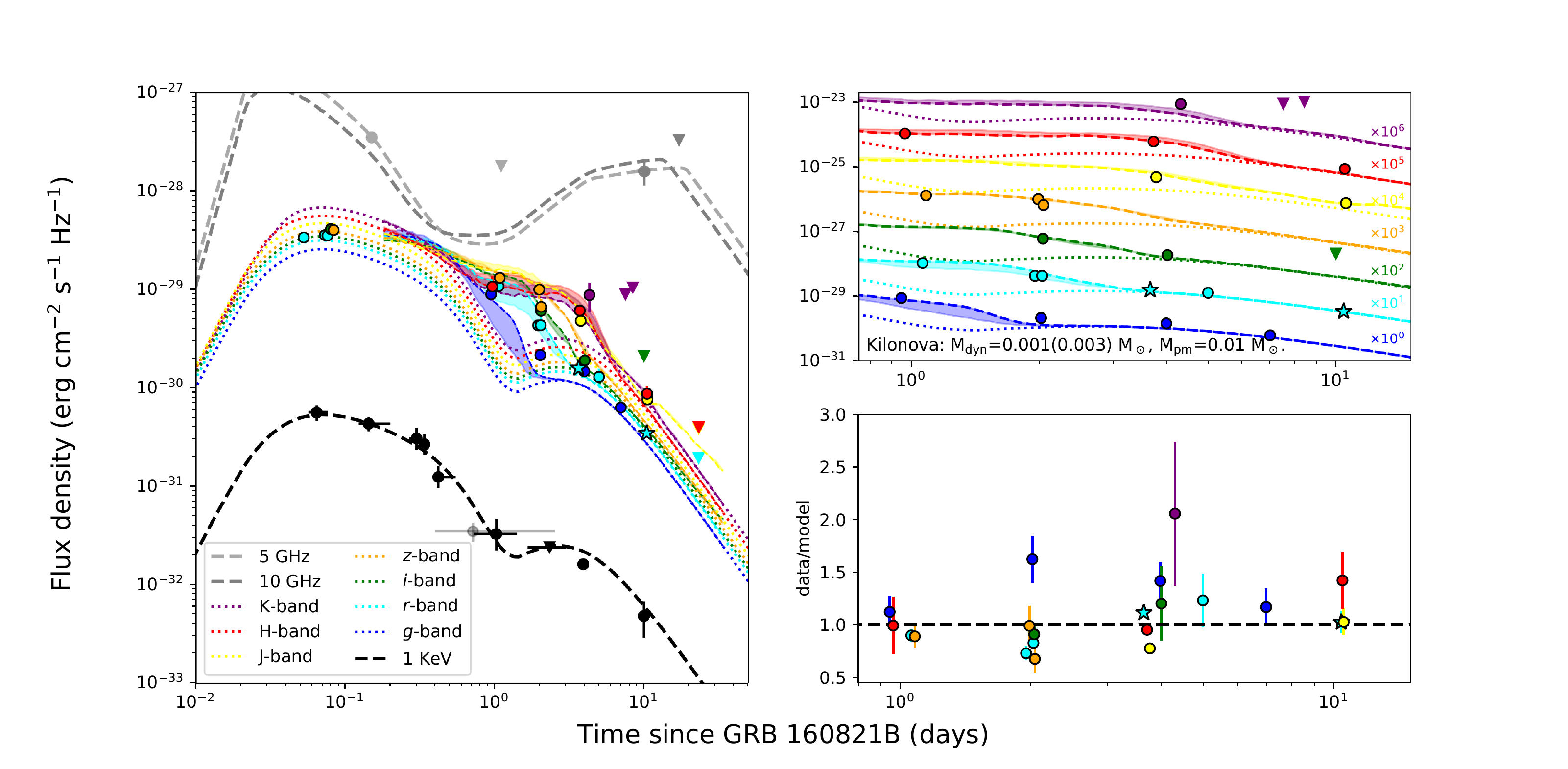}
\caption{Left panel: X-ray, optical, near-infrared, and radio frequency observations of \grb{} afterglow.
{Star markers in the $r$-band indicate HST/WFC/F606W data points}.
Errors are 1$\sigma$ and upper-limits are shown as triangles.
Overplotted are the afterglow lightcurves for a two episode jet and afterglow plus kilonova lightcurves between 0.1 and 30 days, as described in the text.
The reverse shock is dominant at $\lesssim1$ day at 5 GHz (light grey dashed line). 
The re-binned \swift/XRT and \xmm{} data (black markers) show the complex behaviour of the afterglow indicative of a two episode outflow. 
The optical data are clearly in excess above the afterglow model (dotted lines) in blue to red between $\sim 1$ and 5 days.
{The afterglow plus the preferred kilonova model are shown as coloured dashed lines where the shaded region indicates the parameter space for a dynamical mass in a range $0.001 - 0.003$ M$_\odot$, where a higher dynamical mass reduces the $g$-band flux at $\sim1$\,day and increases the K-band flux at $\sim4$\,days.}
Top right: Zoom plot of the optical to near-IR.
{The in-band light-curves are separated by the factor indicated for each line.} 
Afterglow is shown as dotted lines and the sum of the afterglow model and the kilonova model is shown as dashed lines. 
Bottom right: the residual of the best fitting afterglow plus kilonova model and the data.
The line and marker colours for each band are given in the legend. \label{fig:lco}}
\end{figure*}

\subsection{At radio wavelengths}
Radio observations show a fading source between $\sim0.1$ and 1 day, but a detection at $\sim10$ days indicates continued radio afterglow emission {as shown by the red contours in Fig. \ref{fig:field} (Note: the small apparent offset between the radio and optical positions is consistent with the effects of noise in the radio map, given the low $S/N$)}. 
The late afterglow is limited by a non-detection at $\sim17$ days.

\subsection{Interpretation}

A kilonova component is likely to peak in the optical within one to two days post-merger,
leading us to expect the $r$-band  flux to be dominated by afterglow at the early ($\sim0.1$\,days) and late ($\sim10$\,days) epochs.
Inspection of the spectral energy distribution at $\sim0.1$ days between the X-ray (1 keV) and the $r$-band optical data reveals $\beta=0.66\pm0.03$, where $F_\nu \propto \nu^{-\beta}$,
and is consistent with $\beta=0.68\pm0.07$ at $\sim10$\,days in agreement with this expectation (see Figure~\ref{fig:lcx}).
Using the broader spectral index limits at $\sim10$ days, and assuming a temporal decline as $F_\nu\propto t^{-\alpha}$, where $\alpha=3(p-1)/4$, the power-law behaviour for the limits on $p$ from $p=2\beta+1$ is shown.
A break in the light-curve at $t_j\sim7$ days is required, where $\alpha=-p$ at $t>t_j$;
this break will be achromatic.
The X-ray light-curve drops significantly below the lower-limit ($p=2.23$) power-law extrapolated to earlier times from $\sim4$ days.

The X-ray light-curve exhibits an earlier break at $t\sim0.35$ days, and a late-time excess.
Afterglow variability is discussed in \cite{ioka2005}, and such an excess is expected from either a refreshed shock where a slower shell catches up with the initial decelerating outflow \citep[e.g.][]{panaitescu98, zhang02}, or a structured jet with an angle-dependent energy and Lorentz factor distribution \citep[e.g.][]{lamb17}.
{By assuming the jet structures used to model the afterglow to GRB\,170817A in \cite{lamb19}, where on-axis the resultant GRB would have been consistent with the short GRB population \citep[e.g.][]{salafia2019}, then from the observed $\gamma$-ray energy of GRB\,160821B we can estimate the system inclination following \cite{ioka2019}.
For a Gaussian structure with GRB\,170817A-like core energy $[\log_{10}(E_c)=52.4^{+0.4}_{-0.5}]$, then to reproduce the prompt $\gamma$-ray energy of GRB\,160821B, the system should be inclined at $\sim \theta_c + (3\pm2)^\circ$ \citep[see also][]{troja2019_160821B};
for a two-component jet $[\log_{10}(E_c)=52.0^{+0.6}_{-0.9}]$ then the opacity of the low-$\Gamma$ second component must be considered \citep[e.g.][]{lamb16} and the expected inclination would be $\sim \theta_c + (1.5\pm1.5)^\circ$.}
For a structured jet, however, a late-time re-brightening in the afterglow is only expected {for some structure profiles and} at higher inclinations, $\sim (3-5)\times\theta_c$\footnote{{A late excess/re-brightening is not expected from a Gaussian profile structure.}} where bright $\gamma$-ray emission is not expected \citep[see][]{lamb17, lamb2018a, gill2018, beniamini2019, matsumoto2019}.
Considering the bright GRB {we assume that GRB\,160821B is on-axis or very close to on-axis, where the resultant afterglow would behave similarly to the on-axis case regardless of the jet structure \citep[see][]{lamb17}}.
{For our working model} we favour a refreshed shock scenario with two shells where $\Gamma_1>\Gamma_2$, here the subscript indicates the shell order.
If the jet breaks at $t\sim0.35$ days, then the apparent break at $t>4$ days is indicative of a turnover in the light-curve following a significant energy injection episode.

The extended emission at X-ray frequencies lasting until $\sim 200$--300\,s post \grb{} supports continued engine activity beyond the timescale of the GRB.
This X-ray emission is consistent with an outflow episode driven by fallback accretion onto a spinning black hole \citep{rosswog2007, metzger08, nakamura14, kisaka15, yu15, kisaka2017}.
A peak or break time of $\sim4$ days for the refreshed shock indicates that the bulk Lorentz factor of the outflow when the second shell catches the first should be low, with $\Gamma(t)\sim10$ and the second shell will have a Lorentz factor much lower than the value typically expected for a successful GRB, $\Gamma_2 \ll 100$.
Energy dissipated within a low-$\Gamma$ outflow is not expected to be emitted at $\gamma$-ray energies;
$\gamma$-rays injected into the outflow will be coupled to the plasma and these photons will adiabatically cool and thermalise due to scattering.
The effect of these processes is to suppress any resulting emission which will have a spectral peak at $\sim$X-ray frequencies.
Photons that fail to escape from a low-$\Gamma$ jet will be reabsorbed by the outflow and contribute to the jet kinetic energy driving the afterglow \citep{kobayashi01, kobayashi02, lamb16}.
The energy-loss by the photon distribution and re-absorption by the outflow will result in a very low value for the emission efficiency, $\eta$.
This low-$\Gamma$ X-ray extended emission producing shell follows the initial, high-$\Gamma$, GRB producing shell which will decelerate as $\Gamma_1(t)\propto t^{-3/8}$ as it sweeps-up the ambient medium.
However, the second shell encounters very little material and will catch up with the forward shell when $\Gamma_1(t)\sim\Gamma_2/2$ \citep{kumar2000}.
The energy of the second shell refreshes the forward shock resulting in a re-brightening of the afterglow \citep[e.g.][]{granot2003}.

Although limited, the observations at radio frequencies place tight constraints on any possible afterglow, and the afterglow parameters will be constrained by the detection and upper-limits at 1--10 days.
The early radio detection at $\sim 0.1$ days, brighter than the following upper-limits and flux at $\sim10$ days, is likely the result of a reverse shock \citep[e.g.][]{meszaros1997, sari1999, kobayashi2000, kobayashi01, resmi2016, lamb2019b}.
Given the X-ray to optical spectral index $\beta\sim0.66$, the 5\,GHz radio emission at $\sim 0.1$\,days is below the characteristic synchrotron frequency $\nu_m$;
if the $\sim 0.1$\,days radio emission at 5\,GHz belongs to the forward shock, then as $F_{\rm 5 GHz} = F_{\nu, {\rm max}}(\nu_{\rm R}/\nu_m)^{1/3}$ and considering the flux at X-ray frequencies is $F_{\rm X} = F_{\nu, {\rm max}}(\nu_{\rm X}/\nu_m)^{-\beta}$, then $\nu_m \sim 6.4\times10^{14}(F_{\rm X}/F_{\rm 5 GHz})^{\sim1}$ Hz giving $\nu_m \sim 10^{12}$\,Hz.
As $\nu_m \propto t^{-3/2}$ and $t^{-2}$ for the afterglow before and after the jet break, the 5\,GHz radio emission will brighten until a peak when $\nu_m = 5$\,GHz or the jet breaks;
in either case, the upper-limit of $18~\mu$Jy at $\sim1$\,day post \grb\, rules out the earlier detection being due to the forward shock.
This is the first successfully modelled candidate of a reverse shock in an unambiguous sGRB afterglow and indicates that, in some cases, emission from the reverse shock can be bright despite previous non-detections (\citealt{lloyd2018}; however see \citealt{becerra2019} where a reverse shock was recently claimed for the candidate short GRB\,180418A).
Any afterglow model that can explain the behaviour at X-ray frequencies and the early and late optical and near-IR should also be consistent with the detection and limits at radio frequencies.

The afterglow at both radio and X-ray frequencies can constrain the behaviour at optical and near-IR.
These observations indicate an excess in blue at early times followed by a reddening;
this behaviour is indicative of a kilonova.
Previous studies of \grb\ have been restricted to much smaller photometric data-sets and consequently have only drawn weak conclusions about the possibility of a kilonova component and the nature of the afterglow \citep{jin18, kasliwal17b, gompertz18}.
Here, we use the X-ray, early optical and radio constraints on the afterglow emission to interpret the kilonova contribution at optical and near-IR frequencies.
We use the latest kilonova light-curve models based on numerical-relativity simulations to constrain the dynamical and post-merger ejecta masses \citep[e.g.][]{kawaguchi2018}.

\subsection{Afterglow Modelling}

We use the analytic solution for a relativistic blast-wave from \cite{peer2012}, and the method for generating afterglow light-curves from \cite{lamb2018b} to estimate the broadband afterglow for a given set of parameters.
We use the observed data to constrain several of the GRB afterglow parameters.
As the optical flux at $\sim10$ days could still have some kilonova contribution, we use the 1 keV to $r$-band spectral slope at $\sim0.05$ days to estimate $p$, where $\beta\sim0.66$ giving $p=2.3$.
If we assume a prompt efficiency of $\eta\sim0.1$--0.15 \citep{fong2015}, then the isotropic equivalent kinetic energy in the initial outflow is $E_{{\rm k,iso}}\sim(1$-- $2)\times10^{51}$ erg.
{Throughout,} we fix $\varepsilon_B=0.01$ for the forward shock{, consistent with the range for short GRBs \citep{fong2015}}.

The optical flux is approximately flat between 0.05 and 0.07 days; this flatness combined with a likely reverse shock in the radio at the same time indicates that these points coincide with the deceleration timescale for the outflow.
By fixing the ambient density to $n = 10^{-4}$ cm$^{-3}$, consistent with the location in the outskirts of the host galaxy (see Figure \ref{fig:field}), the Lorentz factor of the GRB outflow can be estimated;
$\Gamma_0 \sim 18~ [t_d/(1+z)]^{-3/8} (E_{{\rm k,iso}}/10^{51}~\mathrm{erg})^{1/8}~ (n/10^{-4}~\mathrm{cm}^{-3})^{-1/8} \sim 55$--60, where $t_d \sim 0.06$ days is the deceleration time.
Similarly, the break at $t_j\sim0.35$ days can be used to estimate the jet half-opening angle, $\theta_j \sim 0.05~ [t_j/(1+z)]^{3/8}~(E_{{\rm k,iso}}/10^{51}~\mathrm{erg})^{-1/8}~ (n/10^{-4}~\mathrm{cm}^{-3})^{1/8} \sim 0.033$ rad, or $\sim 1.9$ deg.
{As the break time dominates the opening angle estimation, we can put weak limits on this value of $1.9^{+0.10}_{-0.03}$ degrees (these small errors are only the formal fit uncertainty given this choice of jet model and decomposition of the light curve; the systematic errors from uncertainties in the model assumptions are much greater, and poorly quantifiable), this narrow jet is} consistent with the opening angle range for short GRBs \citep{jin18}.

The forward shock is refreshed at $\sim1$ day, peaking at $\sim3$ days and then declining as $\sim t^{-p}$.
We assume that the second shell has the same half-opening angle as the first.
As the jet has broken, side-ways expansion could widen the initial blast-wave and the second shell will only refresh the blast-wave with an opening angle $\leq \theta_j$.
By assuming that the radius of the blast-wave is roughly constant after the jet break\footnote{The sideways expansion does not halt the radial progress of the jet \citep{granot2012, lamb2018b}; by assuming that it does, we can place a lower-limit on the Lorentz-factor of the second shell.} 
then the Lorentz factor of the second shell is
\begin{equation}
    \Gamma_2 \gtrsim 47.4 \left(\frac{1+z}{t_c}\right)^{1/2}\left(\frac{E_{{\rm k,iso}}}{10^{51}\,\mathrm{erg}}\right)^{1/6}\left(\frac{n}{10^{-4}\,\mathrm{cm}^{-3}}\right)^{-1/6}\theta_j^{1/3},
\end{equation}
where $\Gamma_2\gtrsim16$ for an observed collision time $t_c\sim1$ day.
The Lorentz factor of the forward shock at the collision is then $\Gamma_1(t)\gtrsim8$.

We find that if the forward shock is refreshed when $\Gamma_1(t)=12$ and the resulting blast-wave has $12.5\times E_{{\rm k,iso}}$ of the initial outflow energy then the afterglow can account for the X-ray excess at $\sim 4$ days.
The radio afterglow at $\sim 10$ days constrains the micro-physical parameter $\varepsilon_e\sim0.3$, so as not to overproduce the radio flux.
We assume throughout that the initial and final blast-wave have identical micro-physical parameters $\varepsilon_B$ and $\varepsilon_e$, electron index $p$, and $\theta_j$.

The early radio point at $\sim 0.1$ days requires a significant reverse shock.
For this point to be forward shock dominated the X-ray and optical data constrain the characteristic synchrotron frequency to $\nu_m\sim10^{12}$\,Hz, much lower than the model estimate of $\nu_m\sim3.5\times10^{14}$\,Hz.  
As $\nu_m \propto \Gamma^4 \varepsilon_B^{1/2} n^{1/2} \varepsilon_e^2$, then the parameters that can successfully explain the X-ray and optical afterglow would need significantly lower values.
Such lowered parameter values result in an afterglow that is inconsistent with the other observations and unphysical parameters in many cases.
Following \cite{harrison2013}, the characteristic synchrotron frequency $\nu_m$ and the maximum flux $F_{\nu,{\rm max}}$ for the reverse shock can be found from the forward shock parameters.
The reverse shock flux before and after the peak will scale following \cite{kobayashi2000};
for the thin shell case and our parameters, the flux pre-peak will scale as $F_\nu\propto t^{5.7}$ and post peak $F_\nu\propto t^{-2.05}$.
To accommodate the early radio detection, we need to use a magnetization parameter of $R_B\sim8$.
The model light-curve is shown in Figure \ref{fig:lco}, where we have taken an initial kinetic energy of $E_{{\rm k,iso}}=1.3\times10^{51}$ erg and $\theta_j=0.033$, with all other parameters as discussed.

\subsection{The Kilonova Modelling}

The kilonova appears as an excess in the optical above the afterglow.
From Figure \ref{fig:lco}, where the optical afterglow is shown as dotted lines, it is clear that all bands are in excess at $\sim 1$ day post-burst.
The bluer bands ($g$, $r$, and $i$) follow the afterglow from $\sim5$ days whilst the redder bands ($J$, $H$, and $K$) remain in excess until $\sim 10$ days post GRB.

Using two-component kilonova models from \cite{kawaguchi2018}{, $K$-corrected to $z=0.16$,} we find the model parameters via a {$\chi^2$ minimisation} fit to the data for the kilonova plus model afterglow.
The kilonova is best described\footnote{{The models have masses drawn from the parameter-grid $M_{\rm {dyn}}=[0.001,~0.002,~0.003,~0.005,~0.01]$\,M$_{\odot}$, and $M_{\rm{pm}}=[0.01,~0.02,~0.03,~0.05,~0.1]$\,M$_{\odot}$}.} by a secular ejecta (or post-merger wind {driven by viscous and neutrino heating})  with a mass $M_{\rm pm} = 0.01$ M$_\odot$, and a dynamic ejecta mass $M_{\rm dyn} = 0.001$\,M$_\odot$.
The density profile for each ejecta component is given by
\begin{equation}
    \rho(r,t) \propto \left\{ \begin{array}{lr}
    r^{-3}~t^{-3} &  0.025c\leq r/t\leq 0.15c,\\
    r^{-6}~\zeta(\theta)~t^{-3} &  0.15c\leq r/t\leq0.9c.
    \end{array}
    \right.
\end{equation}
Here the top condition is for the secular ejecta, and the bottom condition for the dynamic ejecta.
We find good fits for an upper-limit for the secular ejecta velocity, and lower-limits for the dynamic ejecta velocity, of $0.1 - 0.15c$.
The function $\zeta(\theta)$ describes the angular distribution of the dynamic ejecta, and is given by
\begin{equation}
    \zeta(\theta) = 0.01+\frac{0.99}{1+e^{-20(\theta-\pi/4)}},
\end{equation}
where $\theta$ is the angle from the central axis.

The element abundances for the ejecta are determined following the results of r-process nucleosynthesis calculations by \cite{wanajo2014} and assuming that the secular and dynamic ejecta have initially flat electron fraction $Y_e$ distributions ranging from 0.3 to 0.4 and from 0.1 to 0.4, respectively.
Radiative transfer simulations were performed from 0.1 to 30 days resulting in a lightcurve with a statistical error in each band $\sim 0.1 - 0.2$ magnitudes.

The kilonova fit to the data depends on the afterglow subtraction, however, the precise details of the afterglow parameters are not crucial. 
As the optical afterglow is typically in the same spectral regime as the observed X-ray data for sGRBs, and supported by the similar spectral index between optical and X-rays at 0.1 and 10 days, then the optical afterglow will follow that at X-ray frequencies during the kilonova peak.
The X-ray data extrapolated to the optical at $\sim 1- 4$\,days post-burst indicates that the afterglow contributes $\sim10\%$.
The typical photometric uncertainty is $\sim10\%$, and the kilonova model uncertainty is $\sim10\%$.
Combining these uncertainties, and using the analytic scaling for luminosity with mass $L\propto M^{0.35}$ \citep[e.g.][]{grossman14}, we can give limits on the mass estimates from the kilonova model fit of $\sim \pm60\%$, {however, we emphasize that both the masses and the uncertainties are model specific}.

\section{Discussion}
\label{sec:discussion}

We have shown that the afterglow of \grb\ with extended X-ray emission until $\sim 300$\,s post-burst exhibits a reverse shock at early times and a refreshed shock at late times.
Early time observations at radio wavelengths require a reverse shock, while the complex light-curve at X-ray frequencies observed by {\it Swift}/XRT and \xmm, combined with late time radio observations reveal a break at $\sim0.35$\,days and a re-brightening at $>1$ day.
The jet is very narrow, at $\theta_j\sim 1.9$ degrees, and the slower second outflow episode that refreshes the forward shock carries significantly more energy than the initial outflow.
However, the total combined energy of the jets, $E_j\sim0.9\times10^{49}$ erg, is consistent with the short GRB population \citep{fong2015}.

Extended emission can be the result of a magnetar \citep[e.g.][]{fan2006, metzger08, bucciantini2012, gompertz2013, gibson2017}, or energy dissipated within a jet launched due to mass fallback onto the central compact object \citep{fan2005, rosswog2007, kisaka15, kisaka2017};
see also \cite{barkov2011} for a two-component jet model.
The refreshed shock at late times requires a second episode of jet activity and fallback accretion onto the central compact object supports both this late-time re-brightening and the extended emission.
From the afterglow modelling, the second jet episode has a Lorentz factor of $\Gamma_2\sim24$.
Internal energy dissipation within such a low-$\Gamma$ jet is expected to be suppressed due to a large optical depth, see \cite{lamb16}, however, any resulting emission will peak at X-ray frequencies and have a longer timescale than the initial dissipation timescale.
Considering the energy required to refresh the forward shock, the efficiency of energy dissipation within the fallback launched jet is $\eta\sim10^{-3}$, consistent with the expectation from a low-$\Gamma$ outflow \citep{lamb16}.
The fallback mass required to launch such an energetic second outflow can be estimated following \cite{kisaka2017} giving a mass $\sim 2\times10^{-3}$ M$_\odot$.

As well as the EE and the refreshed shock, the afterglow reveals a reverse shock \citep[the first confirmed reverse shock in an sGRB, see][who highlight the lack of observed reverse shocks in sGRBs]{lloyd2018};
such a shock propagates into the colder and denser inner shell.
To recreate the reverse shock emission we follow \cite{lamb2019b} and require a magnetization parameter of $R_B\sim8$.
Thus the magnetic field within the shell is much larger than the magnetic field induced by the forward shock.
A high magnetic field indicates that the shell is endowed with primordial magnetic fields from the central engine.

\begin{figure}
\plotone{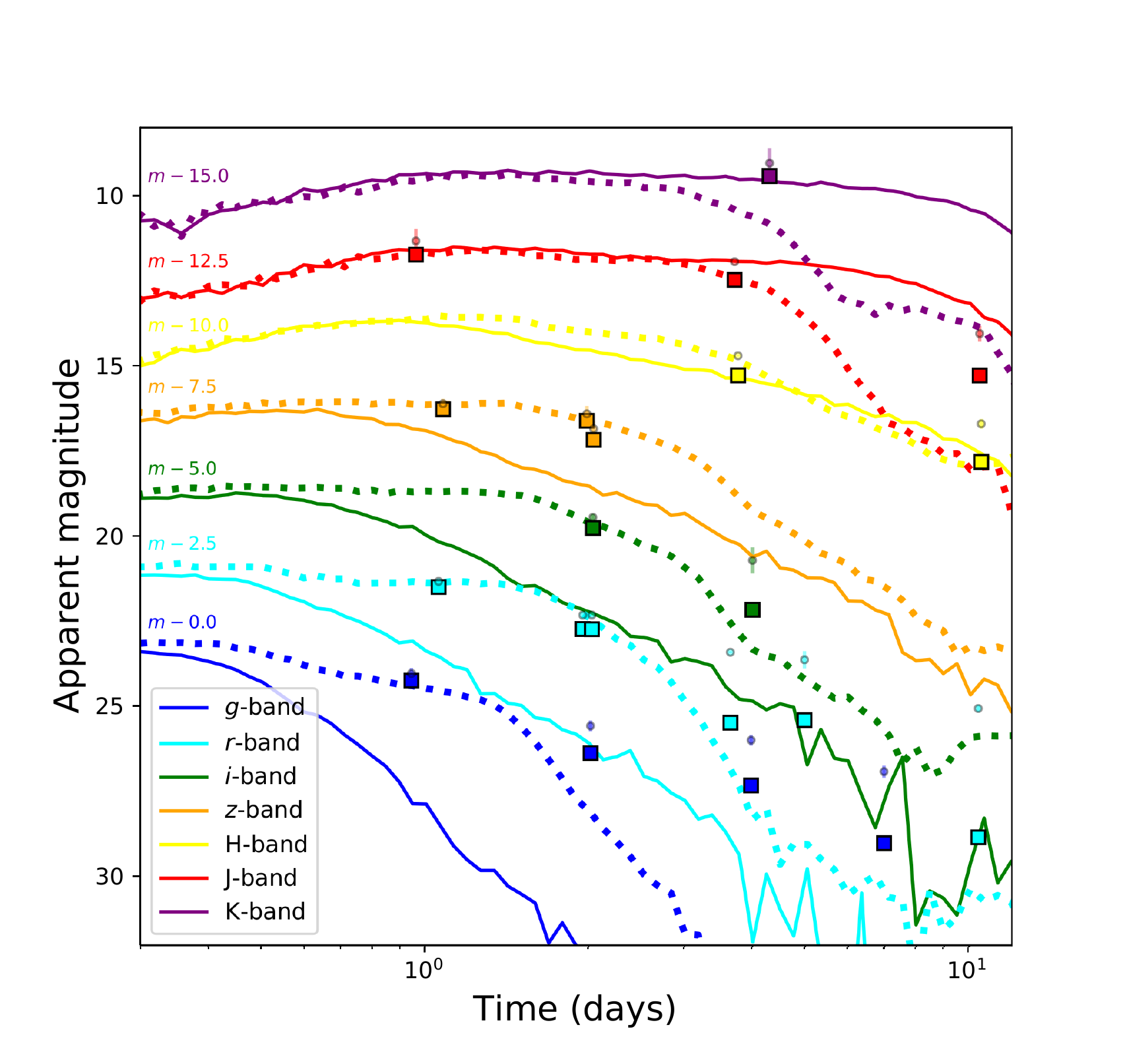}
\caption{Kilonova model light-curves for a BNS to a short-lived hyper-massive neutron-star (the model used by our analysis) is shown as dotted lines, compared to the scenario where the BNS promptly forms a black hole, shown as solid lines.
{The in-band flux has been separated by a factor, annotated on each light-curve.
The square markers show the data with the model afterglow flux removed, the original data are shown with error bars and a small circle.}
The prompt collapse scenario {under-produces the early, $\lesssim4$\,days,} bluer observations. 
}
\label{fig:rvsb}
\end{figure}

In addition to these afterglow features, a kilonova is present at optical and near-IR frequencies.
The best fitting model is one represented by a dynamic ejecta mass of $\sim0.001$ M$_\odot$ and a secular ejecta mass $\sim0.01$ M$_\odot$.
The secular ejecta mass, required for the early blue excess, is consistent with the expectation of the mass-loss from a torus surrounding a massive neutron star \citep{fujibayashi2018, fernandez2019}.
However, {the best-fit model from our parameter sample under-predicts the observed $g$-band emission at $\sim2$ and $\sim4$\,days post-burst, this is likely due to the finite parameter spacing of the kilonova model samples.}
A small secular ejecta mass $\sim0.01$ M$_\odot$ and the low dynamic ejecta mass $\sim0.001$ M$_\odot$ may indicate that the remnant collapses to a black hole promptly after the merger \citep{kiuchi2009, sekiguchi2016, coughlin2018, radice2018}.
In such a scenario the electron fraction, $Y_e$, will be lower.
To test this, we compared the kilonova light-curve of the best-fit model with a model using a lower electron fraction distribution for the post-merger wind $Y_e=0.1-0.3$ as expected from a prompt collapse scenario.
A comparison of the light-curves for these two scenarios was performed, the results indicate that the prompt collapse to a black hole, with a low-$Y_e$ and a higher velocity, will overproduce the red excess at late times and underproduce the early blue excess;
see Figure \ref{fig:rvsb}.
Thus, the observed blue emission in the early phase suggests the existence of a low opacity component, when interpreted as kilonova emission, and we can conclude that a very prompt collapse to a black hole is unlikely to explain the observed transient when considering the observed features.
{Note that the afterglow subtracted data at $\gtrsim4$\,days is typically brighter than the kilonova model we use, especially at K-, J-, $r$- and $g$-bands.
This excess at bluer wavelengths is due to the afterglow subtraction, where the emission is afterglow dominated and the model afterglow slightly under-predicting the observed flux. The observed K- and J-band excesses ($\sim4$ and $\sim10$\,days post-burst) have large associated errors, and the best-fit model is within 2$\sigma$ of each detection without considering the model uncertainty (see Fig.\,\ref{fig:lco}). 
}

Of the five widely discussed GRBs with candidate kilonova contributions to their light-curves -- GRBs 050709, 060614, 070809, 130603B and 150101B \citep{jin16, yang15, jin2019, gompertz18, troja2018} -- the kilonova in \grb\ is the best sampled.
At $\sim0.011$ M$_\odot$, the kilonova in \grb\ has an ejecta mass towards the lower end of the range proposed for any of these other cases, and is
consistent with the $<0.03$ M$_\odot$ found by \cite{kasliwal17b}.
The kilonova following GW170817 had an ejecta mass $\sim0.03-0.05$ M$_\odot$ \citep[e.g.][]{smartt17, pian17}, similar to the mass estimates for sGRB 130603B, $\sim0.03$ M$_\odot$ \citep[e.g.][]{jin16}, whereas, GRB 050709, 060614, 070809 and 150101B have masses $\sim0.05$, $0.13$, $0.015$ and $<0.004$ M$_\odot$ respectively \citep{jin16, yang15}.
However, we note that upper-limits implied by kilonova non-detection in some other sGRBs could indicate the existence of fainter kilonovae indicating still lower ejecta masses\footnote{The heating rates and therefore the estimated
masses depend on the chosen nuclear mass formula \citep[e.g.][]{barnes2016, rosswog2017}.
For the very low $Y_e$ ejecta the r-process path passes close to the neutron-dripline in the nuclear chart, this is experimentally uncharted territory, and we rely on purely theoretical mass formulae.
The amounts of trans-lead nuclei, important since they are efficient in releasing energy and their decay products are efficiently thermalizing with the ambient medium, depend quite sensitively on the chosen mass formula} \citep[e.g.][]{gompertz18}.

The best-fit kilonova model is consistent with the scenario where, following the merger, a massive neutron star survives for a short period \citep{fujibayashi2018}.
This scenario is similar to the case of GRB\,170817A, for which various arguments point to a short-lived massive neutron star \citep[e.g.][see \citealt{piro2019} for an alternative interpretation]{margalit2017, ai2018, pooley2018};
however, the lower ejecta mass in \grb{} could point to a more rapid collapse of the remnant massive neutron star.
Extended emission was present in \grb{} and is used to argue for significant mass fallback in this case, however, for GRB\,170817A \swift/XRT did not begin observations until $\sim15$ hours after the initial burst \citep{evans17} and any EE would have long faded.
The total energy in the jets in \grb\ is lower than the energy required to drive the afterglow to GRB\,170817A and, additionally, the required outflow structure is very different \citep[e.g.][]{lamb19}.
These differences, combined with the lower mass of the ejecta in \grb\ when compared to GRB\,170817A, could offer some clue as to the dynamical differences between mergers and sGRB phenomena.
Understanding these differences may help explain the diversity in sGRB properties; 
especially amongst systems with a similar progenitor i.e. BNS mergers.

\section{Conclusions}
\label{sec:conclusions}

We have reported ground- and space-based optical and near-infrared monitoring of \grb. 
We see clear evidence for red to blue evolution in the colour of the transient, indicative of a kilonova.
The data-set presented here makes the kilonova in \grb\ the best-sampled kilonova without a coincident gravitational wave signal.
We find that a kilonova model with a dynamic ejecta mass $M_{\rm dyn}\sim0.001$ M$_\odot$, a velocity distribution $(0.15-0.9)c$, and a flat electron fraction distribution $Y_e=0.1-0.4$;
and a secular ejecta with $M_{\rm pm}\sim0.01$ M$_\odot$, a velocity distribution $(0.025-0.15)c$, and $Y_e=0.3-0.4$ can best explain the observed emission, while the mass estimates have $\sim60\%$ uncertainty. 
The blue excess, the mass of the dynamic and secular ejecta, and the electron fraction supports the existence of a short-lived massive neutron star that does not immediately collapse to a black hole.

We have also presented {\em Swift} and {\em XMM-Newton} observations of the event and combining with constraints from VLA radio observations find a complex afterglow with a radio-emitting reverse shock into a magnetised shell and a late-time, broadband, refreshed shock.
The jet is very narrow with $\theta_j\sim1.9$ degrees, and the second episode is significantly more energetic than the first.
We find the prompt and extended emission, plus the early- and late-time re-brightening afterglow to be consistent with multiple accretion episodes onto the central compact object with the second episode consistent with a fallback mass of $\sim0.002$ M$_\odot$.

\acknowledgments

The authors thank the anonymous referee for helpful and constructive comments.
GPL additionally thanks Alice Breeveld, Kunihito Ioka, Geoff Ryan, Graham Wynn, and Tomos Meredith for useful discussions and Yizhong Fan for helpful comments.

Partly based on observations made with the Gran Telescopio Canarias (GTC), installed in the Spanish Observatorio del Roque de los Muchachos of the Instituto de Astrof\'isica de Canarias, in the island of La Palma; and with the Nordic Optical Telescope, operated by the Nordic Optical Telescope Scientific Association at the Observatorio del Roque de los Muchachos (program 51-504); and with the Italian Telescopio Nazionale Galileo (TNG) operated by the Fundaci\'on Galileo Galilei of the INAF (Istituto Nazionale di Astrofisica) at the Spanish Observatorio del Roque de los Muchachos (program A32TAC\_5). The development of CIRCE at GTC was supported by the University of Florida and the National Science Foundation (grant AST-0352664), in collaboration with IUCAA. Based on data from the GTC Public Archive at CAB (INTA-CSIC).

This work has made use of data from the European Space Agency (ESA) mission
{\it Gaia} (\url{https://www.cosmos.esa.int/gaia}), processed by the {\it Gaia}
Data Processing and Analysis Consortium (DPAC,
\url{https://www.cosmos.esa.int/web/gaia/dpac/consortium}). Funding for the DPAC
has been provided by national institutions, in particular the institutions
participating in the {\it Gaia} Multilateral Agreement.

NRT, AJL, KW  \& BG have received funding from the European Research Council (ERC) under the European Union's Horizon 2020 research and innovation programme (grant agreement no 725246, TEDE, PI Levan).

AJL, JDL acknowledge support from STFC via grant ST/P000495/1. 

NRT, GPL acknowledge support from STFC via grant ST/N000757/1. 

EP acknowledges support from grant ASI/INAF I/088/06/0

JH was supported by a VILLUM FONDEN Investigator grant (project number 16599).

DBM acknowledges support from the Instrument center for Danish astrophysics (IDA).

The National Radio Astronomy Observatory is a facility of the National Science Foundation operated under cooperative agreement by Associated Universities, Inc.
A.C. acknowledges support from the National Science Foundation CAREER award \#1455090.

AdUP, CCT, ZC, LI and DAK acknowledge support from the Spanish research projects AYA2014-58381-P and AYA2017-89384-P, from the State Agency for Research of the Spanish MCIU through the ``Center of Excellence Severo Ochoa" award for the Instituto de Astrof\'isica de Andaluc\'ia (SEV-2017-0709). AdUP and CCT acknowledge support from Ram\'on y Cajal fellowships (RyC-2012-09975, and RyC-2012-09984). ZC, LI and DAK acknowledge support from Juan de la Cierva Incorporaci\'on fellowships (JdCI-2014-21669, IJCI-2016-30940, and IJCI-2015-26153).

PAE and KLP acknowledge support from the UK Space Agency.

KEH and PJ acknowledge support by a Project Grant (162948--051) from The Icelandic Research Fund.

SR has been supported by the Swedish Research Council (VR) under grant number 2016-03657\_3, by the Swedish National Space Board under grant number Dnr 107/16 and by the research environment grant ``Gravitational Radiation and Electromagnetic Astrophysical Transients (GREAT)" funded by the Swedish Research council (VR) under Dnr 2016- 06012.

%

\vspace{5mm}
\facilities{GTC(OSIRIS/CIRCE), HST(WFC3), NOT(ALFOSC), Swift(XRT and BAT), TNG(DOLORES), VLA, WHT(ACAM)}

\end{document}